%% file: main.tex
\documentclass[a4paper]{PoS}

\usepackage{textcomp} \usepackage{amsfonts} 
\usepackage{amssymb} 
\usepackage{amsmath} 
\usepackage{slashed}
\usepackage{graphicx} 
\usepackage{pifont} 
\usepackage{multirow,lscape}
\usepackage{array} 
\usepackage{longtable}
\usepackage[FIGBOTCAP,TABBOTCAP,tight]{subfigure}
\usepackage{verbatim} 
\usepackage{bm} 
\usepackage{comment}
\usepackage{xcolor} 
\usepackage{url} 

\allowdisplaybreaks

\graphicspath{{./figs/}}

\input{macro.tex}

\title{Nonperturbative Renormalization in the RI-SMOM Scheme and
  Gribov Uncertainty in the RI-MOM Scheme for Staggered Bilinears}

\ShortTitle{RI-SMOM and Gribov Uncertainty}

\author{ Weonjong Lee, Jeonghwan Pak, \speaker{Sungwoo Park}\\ Lattice
  Gauge Theory Research Center, CTP, and FPRD, \\ Department of
  Physics and Astronomy, \\ Seoul National University, Seoul 08826,
  South Korea\\ E-mail: \email{wlee@snu.ac.kr} }

\author{ Jangho Kim\\ National Institute of Supercomputing and
  Networking, \\ Korea Institute of Science and Technology
  Information\\ Daejeon, 34141, South Korea\\ E-mail:
  \email{fraise36@hanmail.net} }

\author{SWME Collaboration}

\abstract{ We present results of renormalization factors for bilinear
  operators obtained using the nonperturbative renormalization method
  (NPR) in the RI-SMOM schemes.  The operators are constructed using
  HYP staggered quarks on the MILC asqtad lattice ($N_f=2+1$). We
  compare results in the RI-SMOM schemes with those in the RI-MOM
  scheme for the $V\otimes S$ and $S\otimes S$ operators. Since we use
  Landau gauge fixing, we study the effect of Gribov ambiguity on the
  wave function renormalization $Z_q$ in the RI-MOM scheme. We find
  that the Gribov uncertainty is negligibly small for $Z_q$ in the
  RI-MOM scheme.}

\FullConference{The 33rd International Symposium on Lattice Field
  Theory\\ 14 -18 July 2015\\ Kobe International Conference Center,
  Kobe, Japan}

\begin{document}

\section{Introduction}


In Ref.~\cite{Bailey:2015tba}, the SWME collaboration reported that
there exists 3.4$\sigma$ tension in $\epsK$ (indirect CP violation
parameter in neutral kaons) between the experiment and the theoretical
evaluation directly from the standard model (SM) with the lattice QCD
inputs.
In order to determine $\epsK$ theoretically, we need to know the kaon
bag parameters such as $B_K$ (in the SM) \cite{ Bae:2014sja} and
$B_{2-5}$ \cite{ Jang:2015sla} (in the BSM\footnote{Here, BSM means
  physics beyond the standard model.}).
Here, we need to know the matching factors which convert lattice data
for $B_i$ into the corresponding quantities defined in the $\MSb$
scheme in the continuum.
Here, we use the non-perturbative renormalization (NPR) method to
determine the matching factors in the RI-SMOM scheme
\cite{Sturm:2009kb}.
The results will be compared with those in the RI-MOM scheme
\cite{Kim:2013bta}.
We will also address Gribov ambiguity in NPR \cite{Gribov:1977wm}.

\section{NPR of Staggered Bilinears in the RI-SMOM Scheme}

A general staggered bilinear operator can be written as
\begin{align}
O^{S \otimes F}_i(y) &= \sum_{AB} \overline{\chi}_i (y_A)
\overline{(\gamma_{S} \otimes \xi_{F})}_{AB} U_{i;AB}(y) \chi_i(y_B)
\label{eq:bilinear}
\end{align}
where $\overline{(\gamma_S \otimes \xi_F)}_{AB} = \frac{1}{4}
\tr[\gamma_{A}^{\dagger} \gamma_{S} \gamma_{B} \gamma_{F}^{\dagger}]$
and $\gamma_A =
\gamma_1^{A_1}\gamma_2^{A_2}\gamma_3^{A_3}\gamma_4^{A_4}$.  The
original coordinate is $y_A$ = $2y+A$ where $A, B$ are hypercube
vectors (each element is 0 or 1). $y$ is the hypercube coordinate on
the lattice with its spacing $2a$.  
$S$ and $F$ stand for the spin and taste degree, respectively.
$i$ is the gauge configuration index and it will be
averaged over gauge ensemble when we calculate the correlation
function. $\chi$ and $\overline{\chi}$ are the staggered quark
fields. 
Here, we use the HYP-blocked fat links for $U_\mu$.

We can obtain the amputated Green's function $\wtd{\Lambda}^{S \otimes
  F}_{c_1 c_2}(\wtd{p}_1+\pi_A, \wtd{p}_2+\pi_B)$ for the bilinear
operators by removing the external quark lines as in
Ref.~\cite{Kim:2013bta}.
Here, we use the reduced momentum $\wtd p_1 \in (-\frac{\pi}{2a},
\frac{\pi}{2a}]^{4}$ defined in the reduced Brillouin zone.
For details, refer to Ref.~\cite{Kim:2013bta}.

We define the projected amputated Green's function $\Gamma$ as
\begin{align}
\Gamma^{\alpha \beta}(\wtd{p}_1, \wtd{p}_2) = \sum_{AB} \sum_{c_1 c_2}
      [\wtd{\Lambda}^{\alpha}_{c_1 c_2}(\wtd{p}_1+\pi_A,
        \wtd{p}_2+\pi_B) \hat{\mathbb{P}}^{\beta}_{BA;c_2 c_1}],
\quad
\hat{\mathbb{P}}^{\beta}_{BA;c_{2}c_{1}} = \frac{1}{48}
\overline{\overline{(\gamma^{\dagger}_{S'}\otimes \xi^{\dagger}_{F'})
}}_{BA}\delta_{c_{2}c_{1}}
\end{align}
where $\alpha = (\gamma_{S}\otimes \xi_{F})$, $\beta =
(\gamma_{S'}\otimes \xi_{F'})$, and $\overline{\overline{(\gamma_S
    \otimes \xi_F)}}_{AB} = \frac{1}{16}{\sum_{CD}}(-1)^{A \cdot C}
\overline{(\gamma_S \otimes \xi_F)}_{CD} (-1)^{D \cdot B}$.
%

\subsection{RI-SMOM schemes}
%
In the RI-SMOM renormalization scheme, we use symmetric momentum $\wtd
p_1^2 = \wtd p_2^2 = \wtd{q}^2 $ at the subtraction momentum
$\wtd{q}\equiv \wtd{p}_1-\wtd{p}_2$.
The subtraction scheme is that $\Gamma^{\alpha\beta}_R (\wtd p_1, \wtd
p_2) = \delta_{\alpha\beta}$, where the sub-index $R$ represents the
renormalized quantity.
We define renormalization factors $Z$ by $\Gamma_R^{\alpha \sigma} =
\sum_\beta Z_q^{-1} Z^{\alpha \beta} \Gamma^{\beta \sigma}_B$ where
where the sub-index $B$ represents bare (=unrenormalized) quantity.

Let us consider the conserved vector current.
There are three different projection methods available in this case
\cite{Sturm:2009kb}.
The first choice is the $\textrm{RI-SMOM}_{\gamma_\mu}$ scheme in
which the subtraction scheme is defined as
\begin{align}
  \Gamma^{V\otimes S}_R (\wtd{p}_1, \wtd{p}_2)|_\text{smom} &\equiv
  \frac{1}{4}\sum_\mu \sum_{AB} \sum_{c_1
    c_2}[\wtd{\Lambda}^{V_\mu\otimes S}_{c_1 c_2}(\wtd{p}_1+\pi_A,
    \wtd{p}_2+\pi_B) \hat{\mathbb{P}}^{V_\mu\otimes S}_{BA;c_{2}c_{1}}
  ]_\text{smom} = 1 \,.
\end{align}
The second choice is the RI-SMOM scheme in which the subtraction
scheme is
\begin{align}
  \Gamma^{V\otimes S}_R (\wtd{p}_1, \wtd{p}_2)|_\text{smom} &\equiv
  \frac{1}{\wtd q^2}\sum_\mu \sum_{AB} \sum_{c_1 c_2}[\wtd q_\mu
    \wtd{\Lambda}^{V_\mu\otimes S}_{c_1 c_2}(\wtd{p}_1+\pi_A,
    \wtd{p}_2+\pi_B) \sum_\nu \wtd q_\nu
    \hat{\mathbb{P}}^{V_\nu\otimes S}_{BA;c_{2}c_{1}}]_\text{smom}
  = 1 \,,
  \label{eq:SMOM}
\end{align}
where $\wtd q = \wtd p_1 - \wtd p_2$.
One advantage of this scheme is that its anomalous dimension for $Z_q$
is already known up to the 4-loop level \cite{Sturm:2009kb}.
The third choice is the RI-SMOM-sin scheme in which the subtraction
scheme is defined as
\begin{align}
  \Gamma^{V\otimes S}_R (\wtd{p}_1, \wtd{p}_2)|_\text{smom} &\equiv
  \frac{1}{\hat{q}^2}\sum_\mu \sum_{AB} \sum_{c_1 c_2}[\hat{q}_\mu
    \wtd{\Lambda}^{V_\mu\otimes S}_{c_1 c_2}(\wtd{p}_1+\pi_A,
    \wtd{p}_2+\pi_B) \sum_\nu \hat{q}_\nu
    \hat{\mathbb{P}}^{V_\nu\otimes S}_{BA;c_{2}c_{1}}]_\text{smom}
  = 1 \,,
  \label{eq:SMOM-sin}
\end{align}
where $\hat{q}_\mu \equiv \sin(a\wtd q_\mu)$ and $\hat{q}^2 = \sum_\mu
\hat{q}_\mu^2$.

The conserved current does not receive any renormalization and so
$Z_{V} = 1$.
Hence, $\Gamma^{V\otimes S}_R = Z_q^{-1} Z_V \Gamma_B^{V\otimes S}=1$
leads to $Z_q = \Gamma_B^{V\otimes S}$.
Similarly, another Ward identity $Z_S \cdot Z_m = 1$ leads to the
identity $Z_m = \Gamma_B^{S\otimes S} / \Gamma_B^{V\otimes S}$.
Here, note that the running of $Z_m$ is different between
$\textrm{RI-SMOM}_{\gamma_\mu}$ and (RI-SMOM \& RI-SMOM-sin) schemes
\cite{Almeida:2010ns}.

\subsection{Simulation Details}
\label{sec:SMOM_simulation_details}
We use $N_f=2+1$, $20^3\times 64$ MILC asqtad ensembles ($a\approx
0.12fm$, $am_\ell/am_s = 0.01/0.05$).
Valence quarks are HYP-smeared staggered fermions with
($am_q =$ 0.01, 0.02, 0.03, 0.04, 0.05).
We use 10 gluon configurations with Landau gauge fixing.
\begin{table}[t!]
  \renewcommand{\subfigcapskip}{0.55em}
  \vspace{-7mm}
  \subtable[Simple momenta]{
    \resizebox{0.48\textwidth}{!}{
      \begin{tabular}{c c c  c c }
        \hline\hline $n_1$ & $n_2$ & $(a\wtd{p})^2$ & $(a\wtd{p})^4$ &
        GeV \\ \hline $(1,1,0,0)$ & $(1,0,1,0)$ & 0.1974 & 0.0195 &
        0.7363 \\ $(2,2,0,0)$ & $(2,0,2,0)$ & 0.7896 &0.3117 & 1.4727
        \\ $(3,3,0,0)$ & $(3,0,3,0)$ & 1.7765 &1.5780 & 2.2090
        \\ $(4,4,0,0)$ & $(4,0,4,0)$ & 3.1583 &4.9873 & 2.9454
        \\ $(5,5,0,0)$ & $(5,0,5,0)$ & 4.9348 &12.1761 & 3.6817
        \\ \hline\hline       
      \end{tabular}
    }
    \label{tab:mom_sim}
  } \hfill \subtable[Complicated momenta]{
    \resizebox{0.48\textwidth}{!}{
      \begin{tabular}{c  c c c c c}
        \hline\hline $n_1$ & $n_2$ & $(a\wtd{p})^2$ & $(a\wtd{p})^4$ &
        GeV \\ \hline $(1,2,3,0)$ & $(-2, 3, 1, 0)$ & 1.3817 &0.9546 &
        1.9482\\ $(2,4,2,0)$ & $(-2, 2, 4, 0)$ & 2.3687 &2.8054 &
        2.5508\\ $(1,3,4,0)$ & $(-3, 4, 1, 0)$ & 2.5661 &3.2924 &
        2.6549\\ \hline\hline
      \end{tabular}
      }
    \label{tab:mom_com}
  } \\
  \vspace{-5mm}
  \caption{List of symmetric momenta: $a \wtd p_\mu\equiv\frac{2\pi
    }{L_\mu} n_\mu$ with $L_S^3\times L_T = 20^3\times 64$. $\wtd{p}^2
    = \displaystyle \sum_\mu \wtd{p}_\mu^2$ and $\wtd{p}^4 =
    \displaystyle \sum_\mu \wtd{p}_\mu^4$\,.}
  \label{tab:mom}
\end{table}
We calculate $\Gamma_B^{\mathcal{O}}(m, \wtd p^2)$ with external quark
momenta $\wtd p$ listed in Table~\ref{tab:mom}.
First, we obtain $Z_\mathcal{O}$ at $\mu_1^2 = {\wtd q}^2$.
Second, we use the RG evolution from the scale $\mu_1$ to
the common scale $\mu_0 = 3$\,GeV.
In the RG running, we use the anomalous dimension obtained using the
perturbation theory as in Refs.~\cite{ Almeida:2010ns,
  Chetyrkin:1999pq}.

\subsection{Chiral extrapolation}

%
\begin{figure}[t!]
  \vspace{-10mm}
  \subfigure{
    \includegraphics[width=0.5\textwidth]{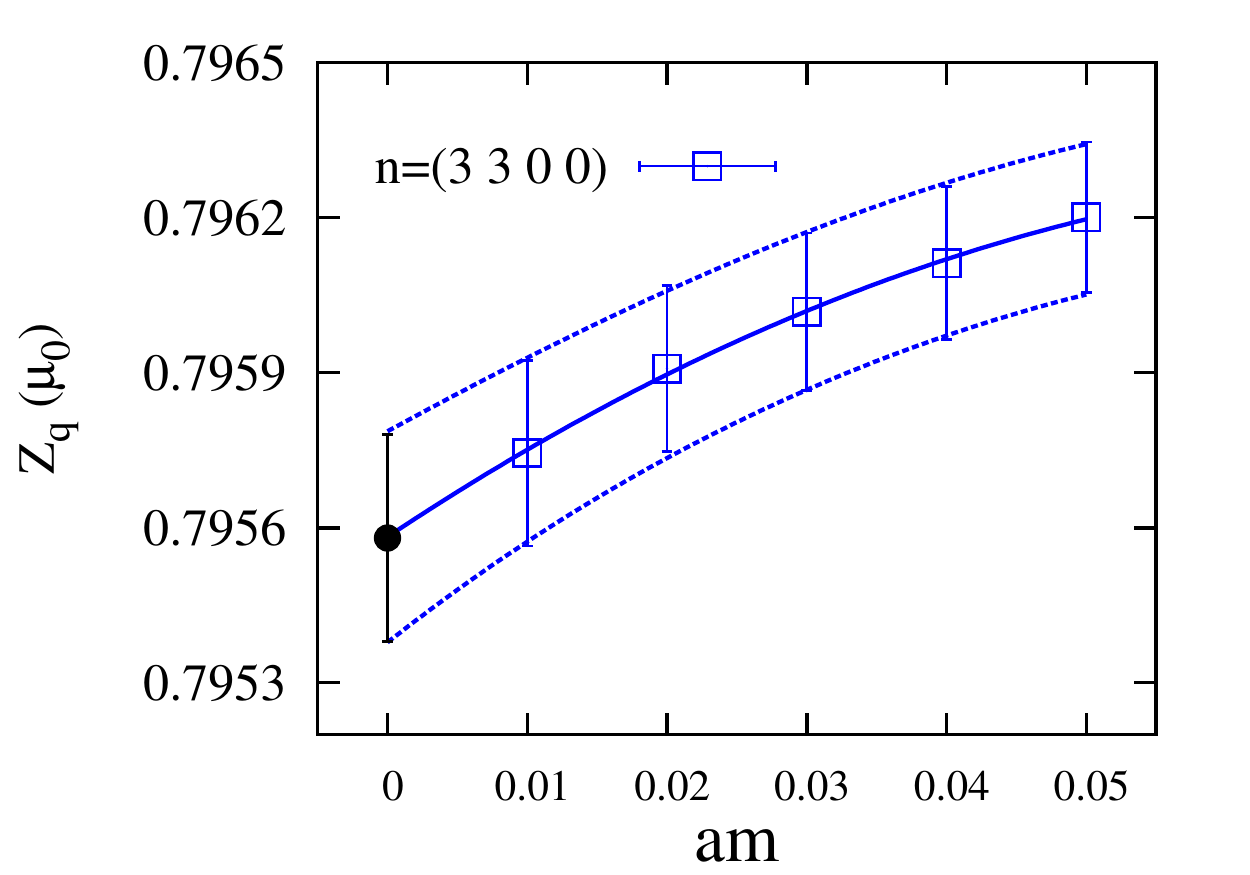}
    \label{fig:m_fit_Z_q}
    }
  \hfill
  \subfigure{ \centering
    \includegraphics[width=0.5\textwidth]{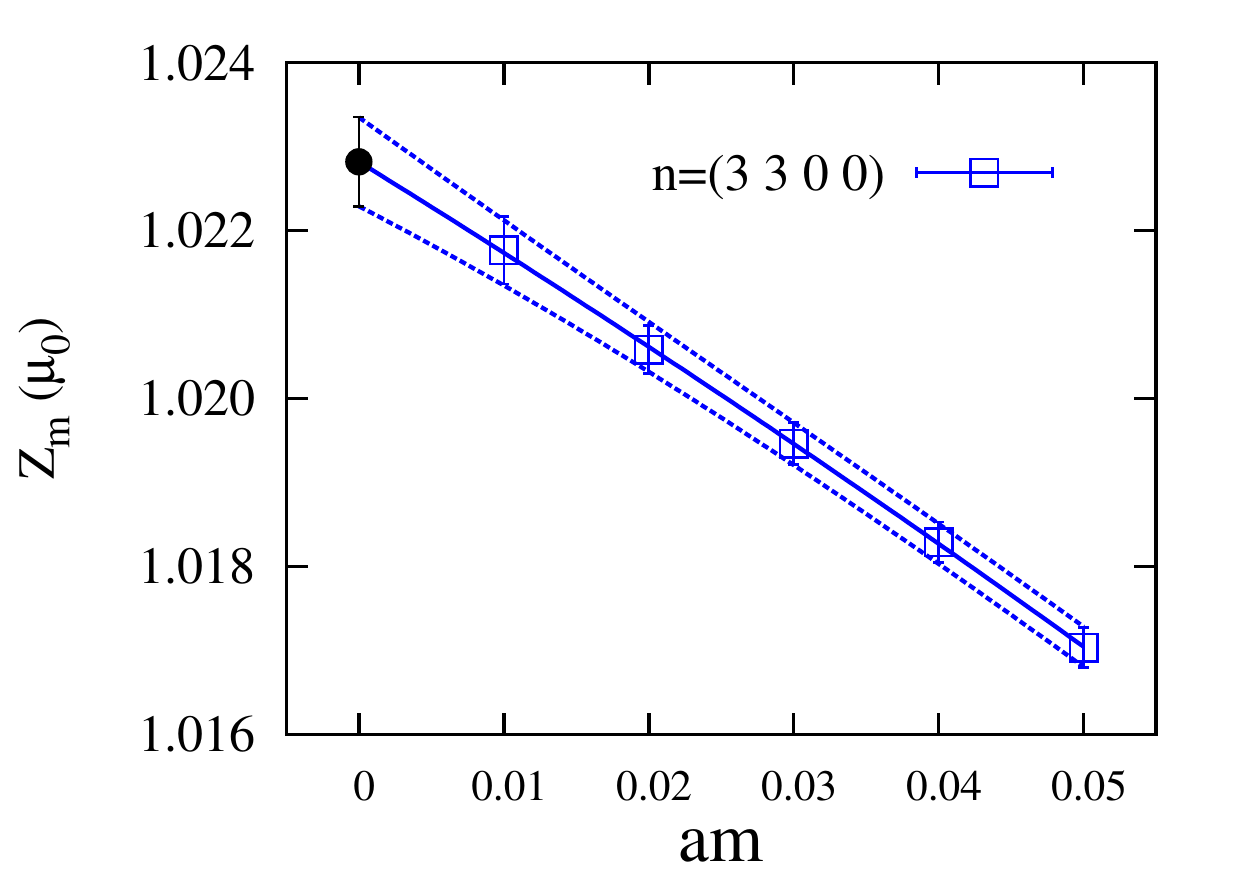}
    \label{fig:m_fit_Z_m}
  }
  \vspace{-7mm}
  \caption{Chiral extrapolation of $Z_q$ in RI-SMOM scheme at
    $\mu_0=3$\;GeV. Black circled points are the chiral limit data obtained from the fitting.}
  \label{fig:m_fit}
\end{figure}

Here, we perform the chiral extrapolation for $Z_q$ and $Z_m$.
In Fig.~\ref{fig:m_fit}, we present results of chiral extrapolation
in $Z_q$ and $Z_m$.
The data in the plots are obtained at the common scale $\mu_0 = 3$\,GeV
with a momentum of (3,3,0,0) in the RI-SMOM scheme.
Here, we use the quadratic fitting to obtain $Z_q$ and $Z_m$ in the
chiral limit.
The fitting results are summarized in Table \ref{tab:m_fit}.

%
\begin{table}[h!]
\center
\begin{tabular}{c | cccc }
\hline
\hline
 & $c_0$ & $c_1$ & $c_2$ & $\chi^2/\text{d.o.f}$ \\
\hline
$Z_q$ & 0.79558(20) & 0.0180(46) & -0.114(52) & 0.0041(45) \\
$Z_m$ & 1.02282(53) & -0.107(18) & -0.18(20) & 0.011(12) \\
\hline
\hline
\end{tabular}
\caption{Fitting result of chiral extrapolation in the
  Fig.~\protect\ref{fig:m_fit} where fitting the function is $c_0 +
  c_1 (am) + c_2(am)^2$.}
\label{tab:m_fit}
\end{table}

\subsection{Results: Momentum Fit for $Z_q$}
\begin{figure}[t!]
  \renewcommand{\subfigcapskip}{-0.55em}
  \vspace{-10mm}
  \subfigure[c-mom]{ 
    \centering
    \includegraphics[width=0.48\textwidth]{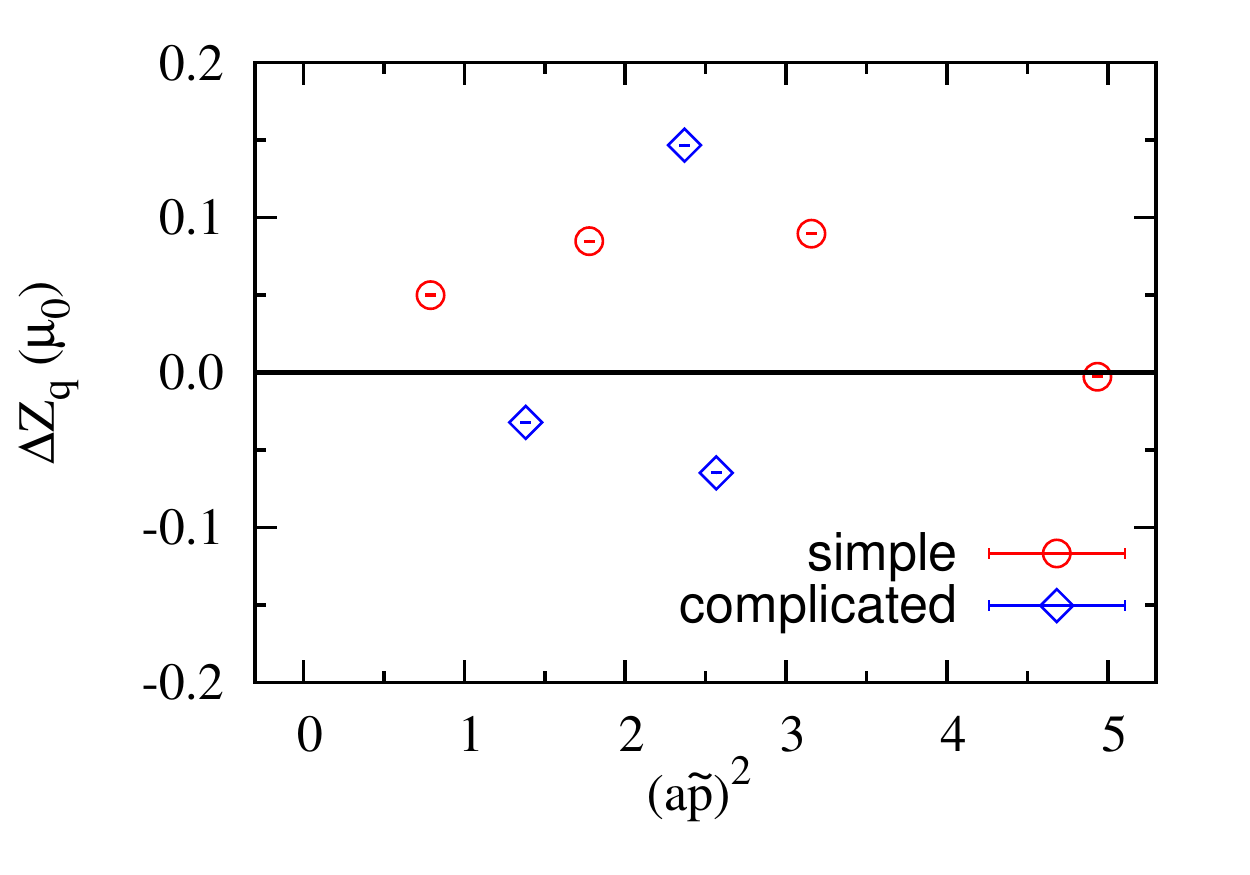}
    \label{fig:c-mom:Z_q}
  }
  \hfill
  \subfigure[s-mom]{
    \centering
    \includegraphics[width=0.48\textwidth]{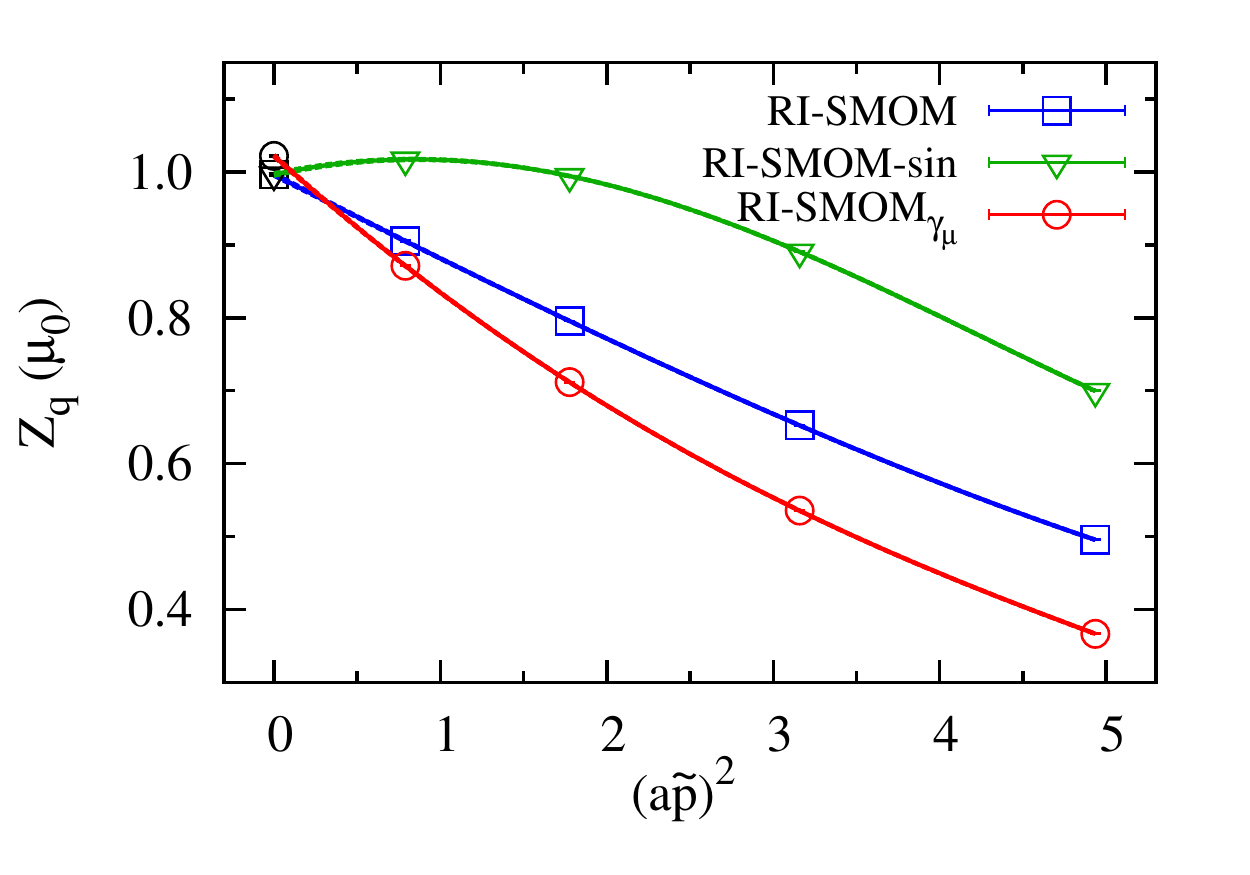}
    \label{fig:s-mom:Z_q}
    }
  \vspace*{-5mm}
  \caption{P-fit results for $Z_q$ in the RI-SMOM schemes:
    \protect\subref{fig:c-mom:Z_q} results of $\Delta Z_q$ with both
    simple and complicated momenta, and \protect\subref{fig:s-mom:Z_q}
    $Z_q$ fits with only simple momenta. Here, c-mom (s-mom)
    represents complicated (simple) momenta.}
  \label{fig:Z_q}
\end{figure}

Here, we explain the p-fit procedure for $Z_q$.
In the case of $Z_q$, we have tried to fit the data of both simple and
complicated momenta to fitting functional forms up to
$\mathcal{O}((a\wtd p)^6)$, and we have failed in finding a reliable
fitting.
In this case, we find typically that $\chi^2/\text{d.o.f} \approx
10^{+6}$.
In Fig.~\ref{fig:Z_q}\;\subref{fig:c-mom:Z_q}, we show $\Delta Z_q =
Z_q(\text{data}) - f(a\wtd p)$ as a function of $(a\wtd p)^2$.
Here, $f(a\wtd p)$ is a trial fitting function.
Large deviation of data points from zero indicates that the fitting
function does not describe the data at all.
Hence, we decide dropping out data of complicated momenta in the
fitting.

We have only 4 data points of simple momenta.
The fitting functional form is
\begin{align}
  f(a\wtd p) &= d_0 + d_1 (a\wtd p)^2 + d_2 ((a\wtd p)^2)^2 + d^b_3
  \left((a\wtd p)^2\right)^3 + d^b_4 \left((a\wtd p)^2\right)^4
\end{align}
Here, note that there is no term like $(a\wtd p)^4$ since it is
not independent of $((a\wtd p)^2)^2$.
First, we fit the data with a fitting function of the first three
terms up to $\mathcal{O}(((a\wtd p)^2)^2)$.
Then, we obtain the fitting scale $\Lambda_n$ using the identity:
$\left( \wtd p^2/\Lambda_n^2 \right)^n = d_n \left( (\wtd p
a)^2\right)^{n}$.
The first trial fit gives $\Lambda_1$ and $\Lambda_2$.
From these values, we find that the minimum bound for $\Lambda_i$ is
$\Lambda \approx 4$\;GeV.
Using this $\Lambda$, we set the Bayesian prior information for the
higher order terms such that $d^b_n = 0 \pm \sigma_n$ with $\sigma_n =
(\Lambda a)^{-2n}$.

For example, on the MILC coarse ($a\approx 0.12$\;fm) ensemble with
$am_\ell/am_s = 0.01/0.05$, the Bayesian prior constraints are
$d^{b}_3= 0 \pm 0.005$ and $d^{b}_4=0\pm 0.0009$.
In Fig.~\ref{fig:Z_q}\;\subref{fig:s-mom:Z_q}, we present the
constrained fitting results for the data set of simple momenta.
We find that results of $Z_q$ in the three RI-SMOM schemes
converge into a point in the limit of $(a\wtd p)^2 = 0$.

%

\subsection{Results: Momentum Fit for $Z_m$}
Results for $Z_m$ are obtained by dividing $\Gamma^{S \otimes S}_B$
by $\Gamma^{V \otimes S}_B$.
Hence, most of lattice artifacts are canceled between the numerator
and denominator, which allows us to fit the data of both simple and
complicated momenta to the fitting functional form:
\begin{align}
  f_{(2)} &= c_1 + c_2 (a\wtd{p})^2 + c_3 (a\wtd{p})^4/(a\wtd{p})^2
  \\
  f_{(4)} &= f_{(2)} + c_4 ((a\wtd{p})^2)^2 + c_5 ((a\wtd{p})^4) + c_6
  ((a\wtd{p})^4/(a\wtd{p})^2)^2 + c_7(a\wtd{p})^6/(a\wtd{p})^2
  \\
  f_{(6)} &= f_{(4)} + c_8 ((a\wtd{p})^2)^3 + c_9 (a\wtd{p})^2(a\wtd{p})^4
  + c_{10} ((a\wtd{p})^4)^2/(a\wtd{p})^2 + c_{11} (a\wtd{p})^6
  \nonumber \\ & \hspace{10mm}
  + c_{12} (a\wtd{p})^4(a\wtd{p})^6/((a\wtd{p})^2)^2
  + c_{13} (a\wtd{p})^8/(a\wtd{p})^2
\end{align}
where the sub-index $n$ of $f_{(n)}$ represents the order
$\mathcal{O}((a\wtd{p})^n)$ of highest order terms included in the
fit.

\begin{figure}[tbhp]
  \renewcommand{\subfigcapskip}{-0.75em}
  \vspace{-5mm}
  \subfigure[$x_m$]{
    \centering
    \includegraphics[width=0.48\textwidth]{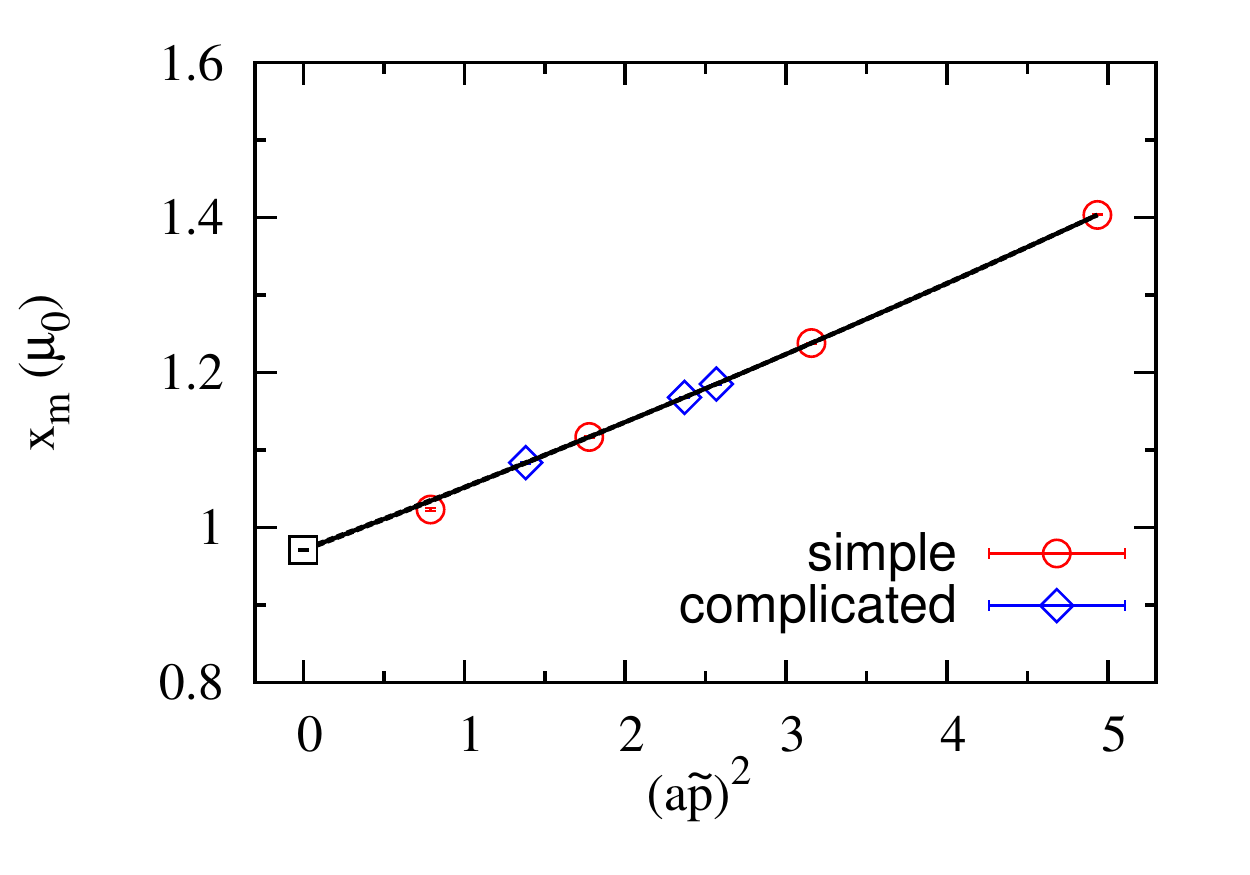}
    \label{fig:Z_m:g_mu:p-fit}
    }
  \hfill
  \subfigure[$\Delta Z_m$]{ 
    \centering
    \includegraphics[width=0.48\textwidth]{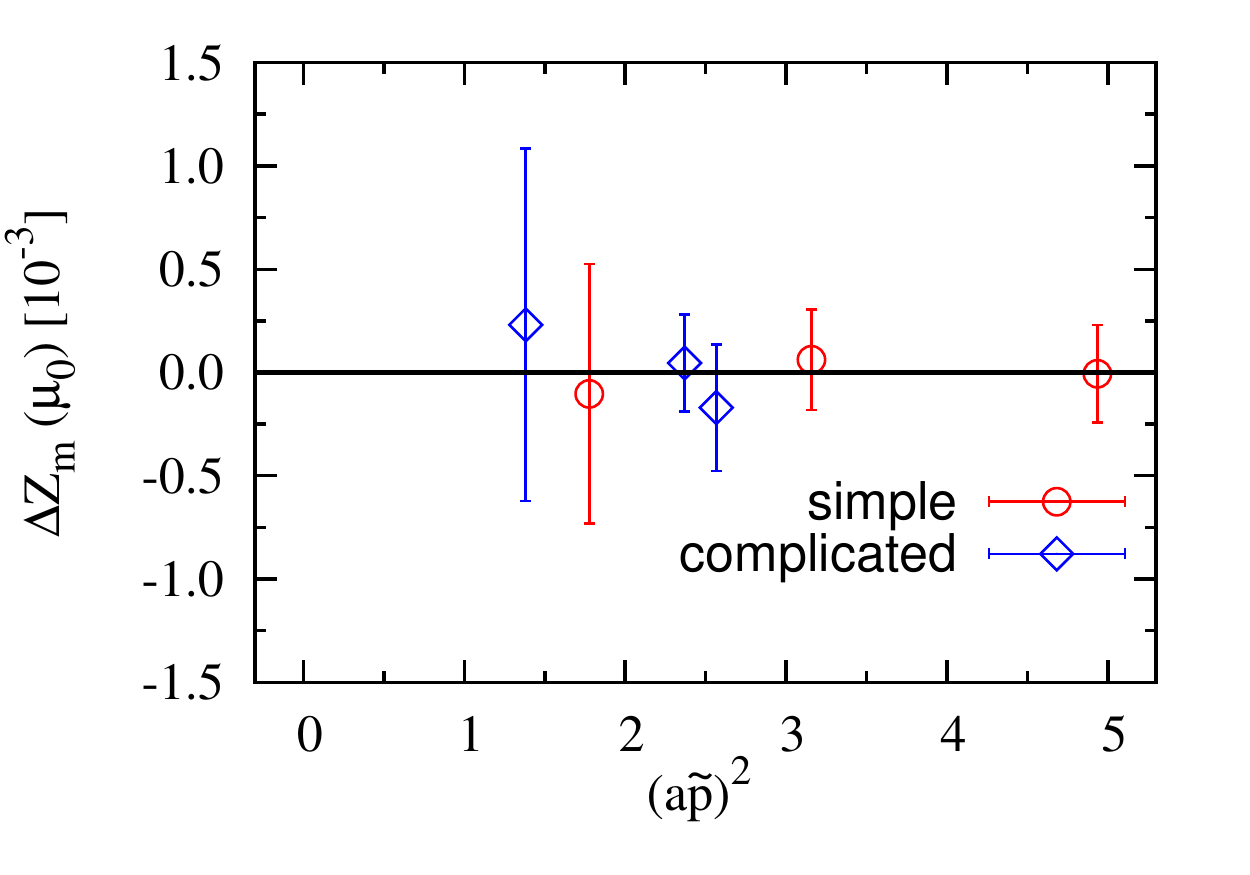}
    \label{fig:Z_m:g_mu:DZ_m}
  }
  \vspace*{-5mm}
  \caption{P-fit results for $Z_m$ in the RI-SMOM$_{\gamma_\mu}$
    scheme: \protect\subref{fig:Z_m:g_mu:p-fit} $x_m$ and
    \protect\subref{fig:Z_m:g_mu:DZ_m} $\Delta Z_m$. }
  \label{fig:Z_m:g_mu}
\end{figure}

In Fig.~\ref{fig:Z_m:g_mu}, we present fitting results for $Z_m$ in
the RI-SMOM$_{\gamma_\mu}$ scheme.
In this fit, we choose $f_{(4)}$ as the fitting function and impose
the Bayesian constraints on $c_{4-7}$: $c_i = 0 \pm \sigma$ and
$\sigma = 1/(a\Lambda)^4$ with $\Lambda = 4$\;GeV for $i =
4,\ldots,7$.
On the MILC coarse lattice, this means that $c_i = 0 \pm 0.03$.
We define $x_m$ as
\begin{align}
  x_m &= Z_m (\text{data}) -\langle c_3 \rangle (a\wtd{p})^4/(a\wtd{p})^2 -
  \langle c_5\rangle(a\wtd{p})^4 - \langle c_6 \rangle
  ((a\wtd{p})^4/(a\wtd{p})^2)^2 - \langle c_7 \rangle
  (a\wtd{p})^6/(a\wtd{p})^2 \;.
\end{align}
Hence, $x_m$ represents $Z_m$ with its lattice artifacts removed and
$\Delta Z_m = Z_m (\text{data}) - f_{(4)}$ corresponds to the fitting
quality.
We present $x_m$ on
Fig.~\ref{fig:Z_m:g_mu}\;\subref{fig:Z_m:g_mu:p-fit}, and $\Delta Z_m$
on Fig.~\ref{fig:Z_m:g_mu}\;\subref{fig:Z_m:g_mu:DZ_m}.
In this fit, $\chi^2/\text{d.o.f.} = 0.20(28)$.
%

\begin{figure}[t!]
  \renewcommand{\subfigcapskip}{-0.75em}
  \vspace{-10mm}
  \subfigure[$y_m$]{
    \centering
    \includegraphics[width=0.48\textwidth]{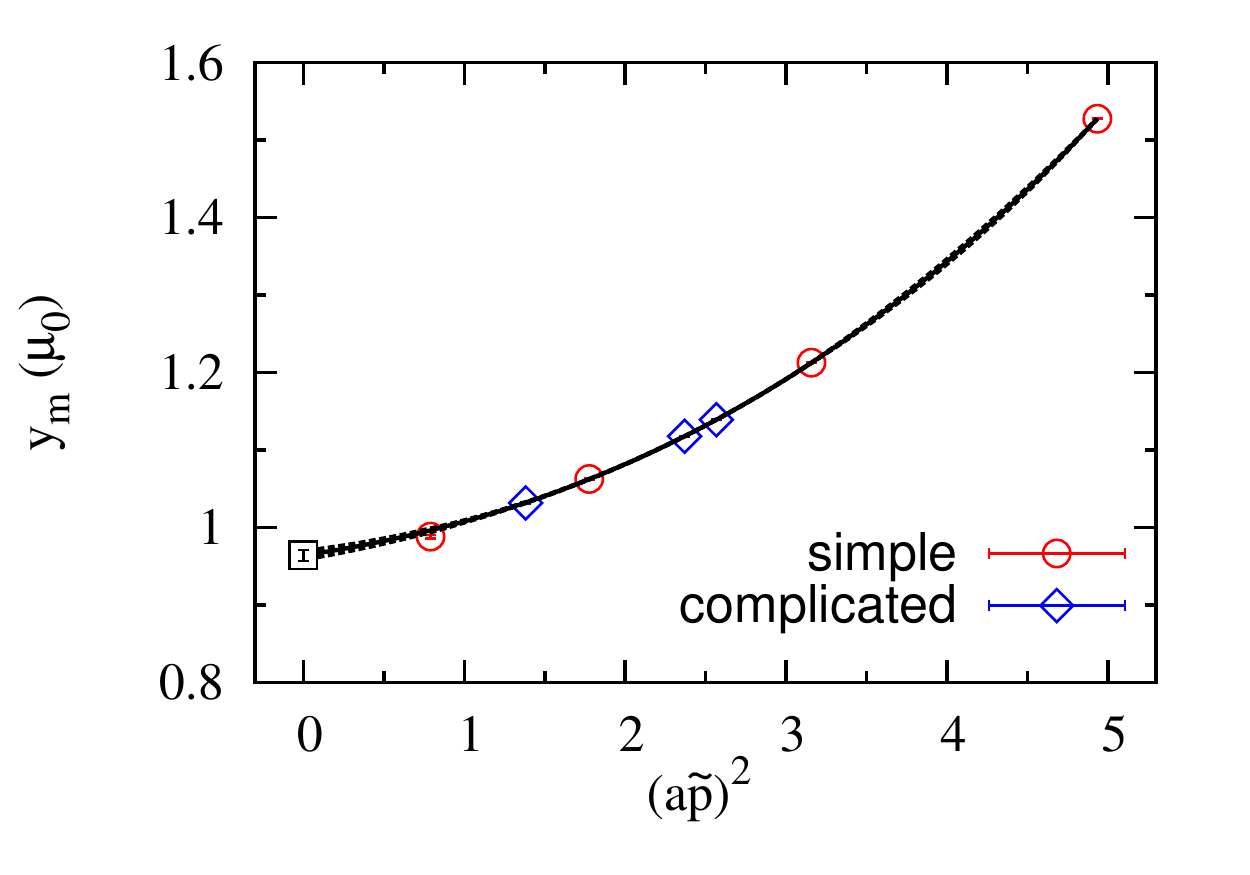}
    \label{fig:Z_m:smom:p-fit}
    }
  \hfill
  \subfigure[$\Delta Z_m$]{ 
    \centering
    \includegraphics[width=0.48\textwidth]{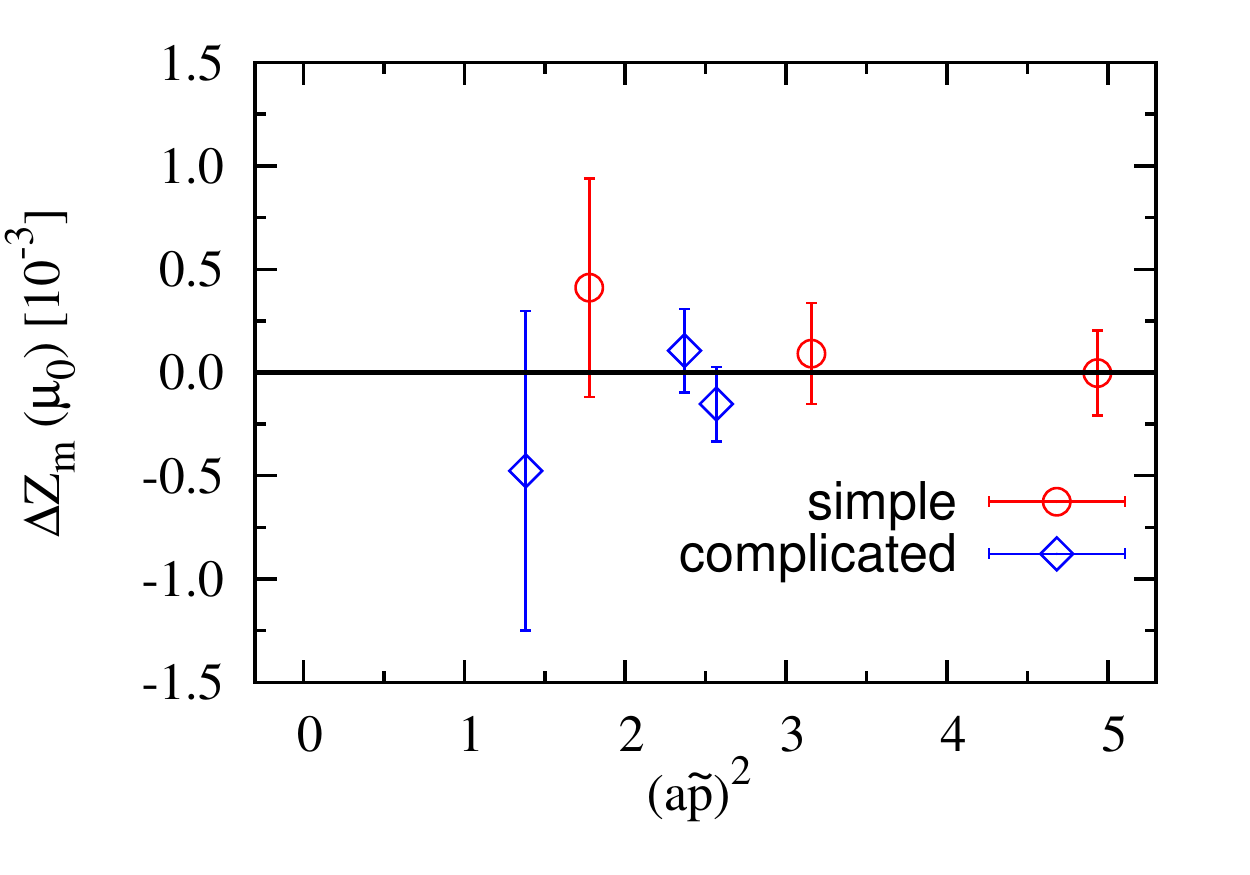}
    \label{fig:Z_m:smom:DZ_m}
  }
  \vspace{-5mm}
  \caption{P-fit results for $Z_m$ in the RI-SMOM scheme:
    \protect\subref{fig:Z_m:smom:p-fit} $y_m$ and
    \protect\subref{fig:Z_m:smom:DZ_m} $\Delta Z_m$. }
  \label{fig:Z_m:smom}
\end{figure}

In Fig.~\ref{fig:Z_m:smom}, we show results for $Z_m$ in the RI-SMOM
scheme.
In this fit, we choose $f_{(6)}$ as the fitting function and impose
the Bayesian prior conditions on $c_{4-13}$.
For $c_{4-6}$, $c_i = 0 \pm \sigma_4$ and $\sigma_4 = 1/(a\Lambda)^4$
with $\Lambda = 4$\;GeV.
For $c_7$, $c_7 = 0 \pm 3\sigma_4$, in order to make the fitting
results consistent with the constraints.
For $c_{8-13}$, $c_j = 0 \pm \sigma_6$ and $\sigma_6 = 1/(a\Lambda)^6$
with $\Lambda = 4$\;GeV.
On the MILC coarse lattice, this means that $\sigma_4 = 0.03$ and
$\sigma_6 = 0.005$.
We define $y_m$ as
\begin{align}
  y_m &= Z_m (\text{data}) -\langle c_3 \rangle (a\wtd{p})^4/(a\wtd{p})^2 -
  \langle c_5\rangle(a\wtd{p})^4 - \langle c_6 \rangle
  ((a\wtd{p})^4/(a\wtd{p})^2)^2 - \langle c_7 \rangle
  (a\wtd{p})^6/(a\wtd{p})^2
  \nonumber \\ & \hspace{5mm}
  - \langle c_9 \rangle
  (a\wtd{p})^2(a\wtd{p})^4 - \langle c_{10} \rangle
  ((a\wtd{p})^4)^2/(a\wtd{p})^2 - \langle c_{11} \rangle
  (a\wtd{p})^6
  \nonumber \\ & \hspace{5mm}
  - \langle c_{12} \rangle
  (a\wtd{p})^4(a\wtd{p})^6/((a\wtd{p})^2)^2 - \langle c_{13}
      \rangle (a\wtd{p})^8/(a\wtd{p})^2
  \;.
\end{align}
Thus, $y_m$ represents $Z_m$ with its lattice artifacts removed.
We also redefine $\Delta Z_m = Z_m (\text{data}) - f_{(6)}$.
We show $y_m$ on Fig.~\ref{fig:Z_m:smom}\;\subref{fig:Z_m:smom:p-fit}
and $\Delta Z_m$ on Fig.~\ref{fig:Z_m:smom}\;\subref{fig:Z_m:smom:DZ_m}.
The fitting quality is $\chi^2/\text{d.o.f.} = 1.15(86)$.
%

In Table \ref{tab:ZqZm}, we summarize our preliminary results for
$Z_q$ and $Z_m$ at $\mu=3$\;GeV in the $\MSb$ scheme.

\begin{table}[!htp]
  \vspace{-5mm}
  \centering
  {
    \begin{tabular}{ l | c | c }
      \hline\hline int.~scheme & $Z_q^{\MSb}(\mu)$  & $Z_m^{\MSb}(\mu)$  \\ 
      \hline 
      $\textrm{RI-SMOM}_{\gamma_\mu}$ & 1.053(1)(15) & 0.920(1)(14)\\ 
      RI-SMOM & 0.984(1)(4) &  0.948(7)(14)\\ 
      RI-SMOM-sin &  0.984(2)(4) & 0.976(7)(15)\\
      \hline
      RI-MOM &  1.060(8)(4) & 0.94(11)(0) \\
      \hline\hline
    \end{tabular}
  }
  \caption{Results of $Z_q$ and $Z_m$ in the $\MSb$ scheme at
    $\mu=3$\;GeV.  They are obtained using the RI-SMOM schemes as an
    intermediate scheme. The first error is purely statistical, and
    the second systematic which comes from the truncation of higher
    order terms in perturbative matching. Here, all the results are
    \textbf{preliminary} in that the error budget is incomplete.}
  \label{tab:ZqZm}
\end{table}

\section{Gribov Uncertainty in RI-MOM}
Landau gauge fixing is done by maximizing the functional $F$: $F = 1/(
2N_c \cdot 4V) \sum_{\mu,x} \textrm{Tr}[U_\mu(x) + U_\mu(x)^\dagger]
$.
%
%
where $N_c=3$, and $V$ is 4-dimensional volume, and $U_\mu$ is
a gluon link field.
In practice, the gauge fixing condition is checked by monitoring
$\theta \equiv 1/(N_c V) \cdot \sum_x \textrm{Tr}
[\Delta(x)\Delta^\dagger(x)]$ such that $\theta < 10^{-14}$.
Here, note that $\Delta(x) \equiv 16 i N_c V \frac{\delta F}{\delta
  \omega^a(x)} T^a$.
We use the Fourier accelerated steepest descent algorithm
\cite{Davies:1987vs} to maximize $F$.
%

It is well known that Landau gauge fixing has Gribov ambiguity
\cite{Gribov:1977wm}: two independent gauge configurations (Gribov
copies) can satisfy the same gauge fixing condition.
In general, we can distinguish different Gribov copies from one another
by monitoring their values of $F$ since $F$ is gauge-dependent.
We start with a mother gauge configuration which has $F = F_m$.
Then we apply randomly gauge transformation to the mother in order to
produce a daughter configuration which has $F = F_d \ne F_m$.
We repeat this procedure 100 times to generate 100 daughter
configurations.
Then, we pick the daughter with $F_d^\text{max}$ which maximize
$\delta F = | F_m -F_d |$.
We measure $Z_q$ on the mother and the daughter with
$F=F_d^\text{max}$.

\begin{figure}[t!]
  \renewcommand{\subfigcapskip}{-0.75em}
  \vspace{-10mm}
  \subfigure[m-fit]{
    \includegraphics[width=0.5\textwidth]{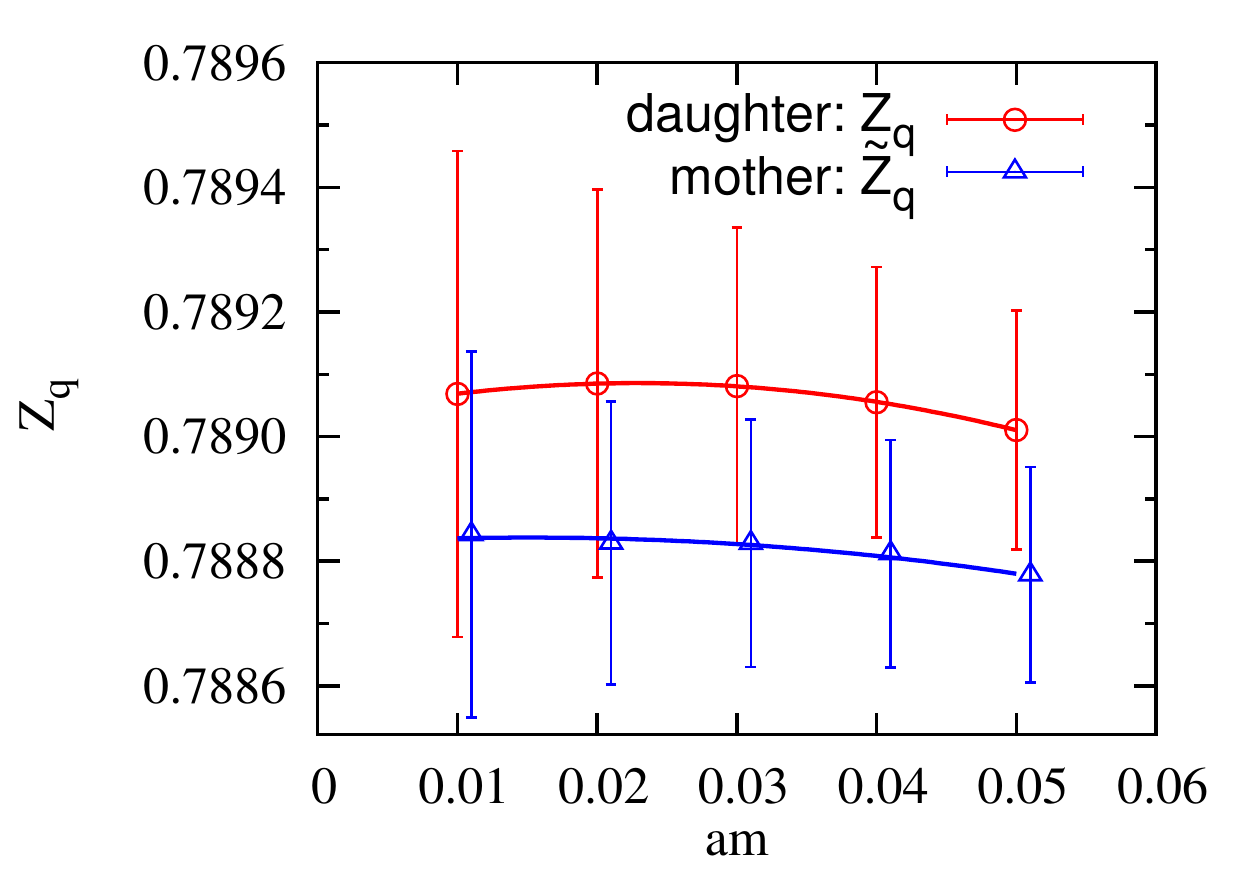}
    \label{fig:gribov:m-fit}
  }
  \hfill
  \subfigure[p-fit]{
    \includegraphics[width=0.5\textwidth]{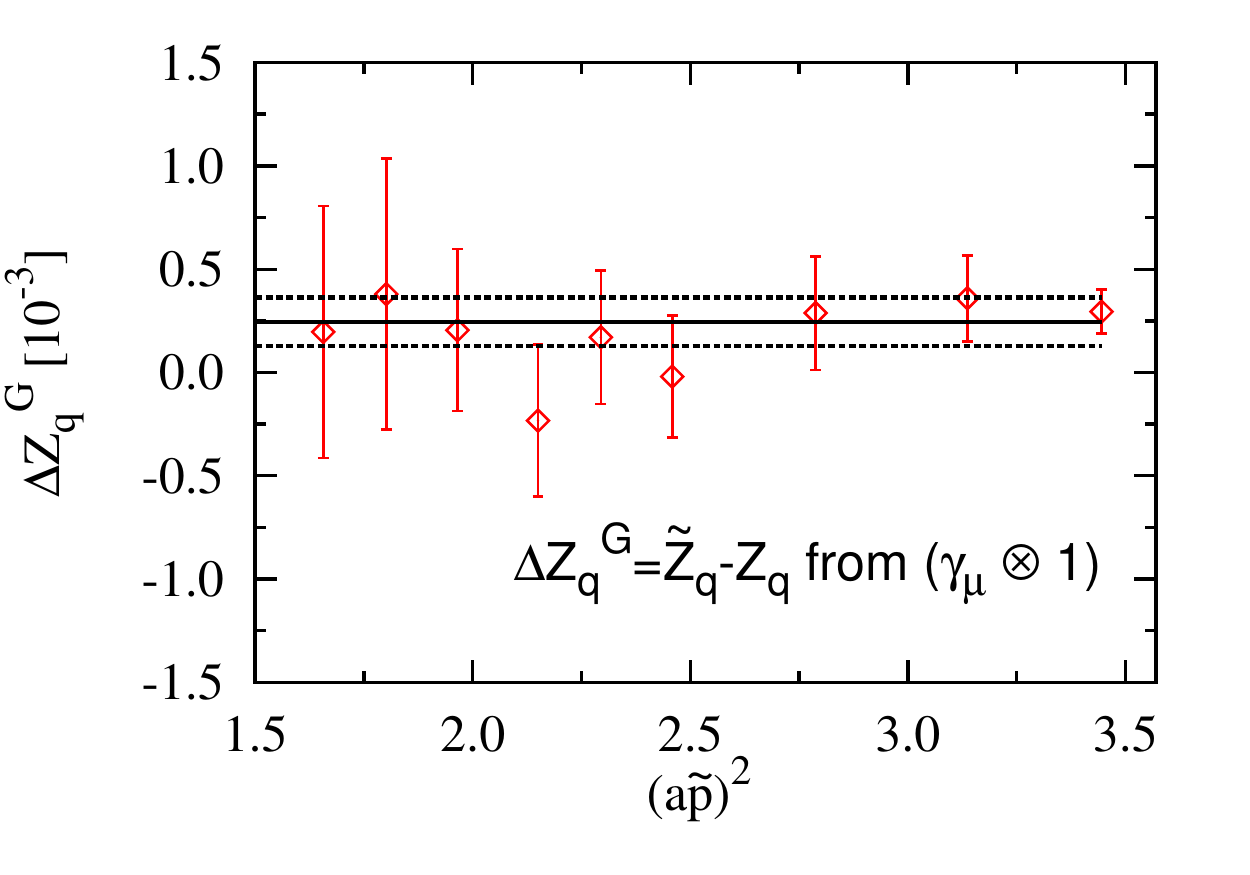}
    \label{fig:gribov:p-fit}
  }
  \vspace{-5mm}
  \caption{ Gribov ambiguity in $Z_q$:
    \protect\subref{fig:gribov:m-fit} m-fit and
    \protect\subref{fig:gribov:p-fit} p-fit of $\Delta Z_q^G=\wtd Z_q
    - Z_q$.  }
  \label{fig:gribov}
\end{figure}

In Fig.~\ref{fig:gribov}, we present results for
$\Delta Z_q^\text{G} = Z_q(\text{daughter}) - Z_q(\text{mother})$.
It turns out that the systematic error due to Gribov ambiguity
is negligibly small ($\approx 0.02\%$).


\acknowledgments
We thank Norman Christ for helpful discussion very much. 
J.~Kim is supported by Young Scientists Fellowship through National
Research Council of Science \& Technology (NST) of Korea.
The research of W.~Lee is supported by the Creative Research
Initiatives Program (No.~2015001776) of the NRF grant funded by the
Korean government (MEST).
W.~Lee would like to acknowledge the support from the KISTI
supercomputing center through the strategic support program
(No.~KSC-2014-G3-003) for the supercomputing application research with
much gratitude.
Part of computations were carried out on the DAVID GPU clusters at
Seoul National University.

\bibliography{refs}


\end{document}

%% file: macro.tex
\newcolumntype{C}[1]{>{\centering\arraybackslash}p{#1}}

\newcommand{\epsK}{\varepsilon_{K}}

\newcommand{\MSb}{{\overline{\text{MS}}}}

\newcommand{\wtd}{\widetilde}
\newcommand{\tr}{\textrm{tr}}